\documentclass[preprint,superscriptaddress,nofootinbib,onecolumn,aps,prd]{revtex4-1}
\pdfoutput=1

\usepackage[utf8]{inputenc}
\usepackage{amsmath,amssymb}
\usepackage{array}

\begin{document}

\title{Double Higgs production at LHC,\\ see-saw type II and
  Georgi-Machacek model}

\author{S.I. Godunov}
\email{sgodunov@itep.ru}
\affiliation{Institute for Theoretical and Experimental Physics, Moscow, 117218, Russia} 
\affiliation{Novosibirsk State University, Novosibirsk, 630090, Russia} 

\author{M.I. Vysotsky}
\email{vysotsky@itep.ru}
\affiliation{Institute for Theoretical and Experimental Physics, Moscow, 117218, Russia} 
\affiliation{Moscow Institute of Physics and Technology, 141700, Dolgoprudny, Moscow 
Region, Russia}
\affiliation{Moscow Engineering Physics Institute, 115409, Moscow, Russia}

\author{E.V. Zhemchugov}
\email{jini.zh@gmail.com}
\affiliation{Institute for Theoretical and Experimental Physics, Moscow, 117218, Russia} 

\begin{abstract}
  The double Higgs production in the models with isospin-triplet
  scalars is studied. It is shown that in the see-saw type II model 
  the mode with an intermediate heavy scalar, $pp\to H+X\to 2h+X$, may
  have the cross section which is compatible with that in the Standard
  Model. In the Georgi-Machacek model this cross section could be much
  larger than in SM since the vacuum expectation value of the triplet
  can be large. 
\end{abstract}

\maketitle

\textit{This paper is our present to Valery Anatolievich Rubakov on his
anniversary. Many students (and not only students) in the world are
studying Physics reading his excellent books, papers and listening his
brilliant lectures.
}

\section{Introduction}
\label{sec:intro}

After the discovery of the Higgs-BE boson at LHC \cite{Higgs2012} the
next steps to check the Standard Model (SM) are: the measurement of
the coupling constants of the Higgs boson with other SM particles
($t\bar t, WW, ZZ, b\bar b,\tau\bar\tau,\dots$) with better accuracy
and the measurement of the Higgs self-coupling which determines the
shape of the Higgs potential. In the SM the triple and quartic Higgs
couplings are predicted in terms of the known Higgs mass and vacuum
expectation value. Deviations from these predictions would mean the
existence of New Physics in the Higgs potential. The triple Higgs
coupling can be measured at LHC in double Higgs production, in which
the gluon fusion dominates: $gg\to hh$. However, the $2h$ production cross
section is very small. According to \cite{deFlorian2014} at
$\sqrt{s}=14~\mbox{TeV}$ the cross section $\sigma^{NNLO}\left(gg\to
  hh\right)=40.2~\mbox{fb}$ with $\left(10-15\right)\%$
accuracy. For the final states with the reasonable signal/background ratios
(such as $hh\to b\bar b\gamma\gamma$) only at HL-LHC with integrated
luminosity $\int{\cal L}dt=3000~\mbox{fb}^{-1}$ double Higgs
production will be found and triple Higgs coupling will be measured
\cite{Baglio2014}\footnote{The decays into $b\bar b\tau\bar\tau$ and
  $b\bar b W^{+}W^{-}$ final states can be even more promising for the
measurement of triple Higgs coupling
\cite{Englert2012,Zurita2012}.}. We are looking for the extensions of
the SM Higgs sector in which the double Higgs production is enhanced. 

One of the well-motivated examples of non-minimal Higgs sector is
provided by the see-saw type II mechanism of the neutrino mass
generation \cite{Schechter1980}. In this mechanism a scalar isotriplet
with hypercharge $Y_{\Delta}=2$ ($\Delta^{++},\Delta^{+},\Delta^{0}$) is added
to the SM. The vacuum expectation value (vev) of the neutral component
$v_{\Delta}$ generates Majorana masses of the left-handed neutrinos. There
are two neutral scalar bosons in the model: the light one in which the
doublet Higgs component dominates and which should be identified with
the particle discovered at LHC ($h; M_{h}=125~\mbox{GeV}$), and the heavy
one in which the triplet Higgs component dominates ($H$). The neutrino
masses equal $f_{i}v_{\Delta}$, where $f_{i}$ ($i=1,2,3$) originates
from Yukawa couplings of Higgs triplet with the lepton doublets. If
neutrinos are light due to a small value of $v_{\Delta}$ while $f_{i}$
are of the order of one, then $H$ decays into the neutrino
pairs. Three states $H^{\pm\pm}$ (or $\Delta^{\pm\pm}$), $H^{\pm}$,
and $H$ are almost degenerate in the model considered in
Sect. \ref{sec:seesaw} and the absence of the same-sign dileptons at
LHC from $H^{\pm\pm}\to l^{\pm}l^{\pm}$ decays provides the lower bound 
$m_{H}>400~\mbox{GeV}$ \cite{HHiggs2012}. We are interested in the
opposite case: $v_{\Delta}$ reaches the maximum allowed value while
neutrinos are light because of small values of $f_{i}$. In this case
$H\to hh$ can be the dominant decay mode of a heavy neutral Higgs. In
this way we get an additional mechanism of the double $h$ production
at LHC.

The bound $m_{H^{++}}>400~\mbox{GeV}$ \cite{HHiggs2012} cannot be
applied now since $H^{\pm\pm}$ mainly decays into the same-sign
diboson \cite{Yagyu1}. We only need $H$ to be heavy enough for $H\to
hh$ decay to occur. This case is analyzed in
Sect. \ref{sec:seesaw}. The invariant mass of additionally
produced $hh$ state peak at $\left(p_{1}+p_{2}\right)^{2}=m_{H}^{2}$
which is a distinctive feature of the proposed mechanism, see also
\cite{Englert2012_2,Barger2014}. 

$H$ contains a small admixture of the isodoublet state which makes gluon
fusion a dominant mechanism of $H$ production at LHC. The admixture of
the isodoublet component in $H$ equals approximately $2v_{\Delta}/v$,
where $v\approx 250~\mbox{GeV}$ is the vacuum expectation value of the
neutral component of isodoublet, and in Sect. \ref{sec:seesaw} for
$\sqrt{s}=14~\mbox{TeV}$ and $M_{H}=300~\mbox{GeV}$ we will get
$\sigma\left(gg\to H\right)\approx 25~\mbox{fb}$. Taking into account
that ${\rm Br}\left(H\to hh\right)$ is about $80\%$, we obtain $50\%$ 
enhancement of double Higgs production in comparison with SM.

Since the nonzero value of $v_{\Delta}$ violates the well checked
equality of the strength of charged and neutral currents at tree
level, 
\begin{equation}
  \label{eq:rho}
  \frac{g^{2}/M_{W}^{2}}{\bar g^{2}/M_{Z}^{2}}=1+2\frac{v_{\Delta}^{2}}{v^{2}},
\end{equation}
$v_{\Delta}$ should be less than $5~\mbox{GeV}$ (see
Sect. \ref{sec:seesaw}). The numerical estimate of $gg\to H$ cross
section was made for maximum allowed value $v_{\Delta}=5~\mbox{GeV}$
when the isodoublet admixture is about $5\%$. 

The bound $v_{\Delta}<5~\mbox{GeV}$ is removed in the
Georgi-Machacek model \cite{Georgi}, in which in addition to
$\vec\Delta$ a scalar isotriplet with $Y=0$ is introduced. If the vev
of the neutral component of this additional field equals $v_{\Delta}$
then we get just one in the r.h.s.~of (\ref{eq:rho}): correction
proportional to $v_{\Delta}^{2}$ is cancelled. Thus $v_{\Delta}$ can be
much larger than $5~\mbox{GeV}$. The bounds on $v_{\Delta}$ come from
the measurement of the $125~\mbox{GeV}$ Higgs boson couplings to
vector bosons and fermions, which would deviate from their SM values:
$c_{i}\to c_{i}\left[1+a_{i}\left(v_{\Delta}/v\right)^{2}\right]$.

The consideration of an enhancement of $2h$ production in GM variant
of see-saw type II model is presented in Sect. \ref{sec:GM}. Since at
the moment the accuracy of the measurement of $c_{i}$ values in $h$
production and decay is poor, $v_{\Delta}$ as large as $50~\mbox{GeV}$
is allowed and $\sigma\left(gg\to H\right)$ can reach
$2~\mbox{pb}$ value which makes it accessible with the integrated
luminosity $\int{\cal L}dt=300~\mbox{fb}^{-1}$ prior to HL-LHC run. We
summarize our results in Conclusions. 

\section{Double $h$ production in $H$ decays at LHC}
\label{sec:seesaw}

\subsection{Scalar sector of the see-saw type II model}
\label{sec:review}

In this subsection we will present the necessary formulas; for a
detailed description see \cite{Accomando}. In addition to the SM
isodoublet field $\Phi$,
\begin{equation}
  \label{eq:Hdoublet}
  \Phi\equiv
  \begin{bmatrix}
    \Phi^{+}\\
    \Phi^{0}
  \end{bmatrix}\equiv
  \begin{bmatrix}
    \Phi^{+}\\
    \frac{1}{\sqrt{2}}\left(v+\varphi+i\chi\right)
  \end{bmatrix},
\end{equation}
in see-saw type II an isotriplet is introduced:
\begin{eqnarray}
  \label{eq:deltatriplet}
  \Delta\equiv\frac{\vec\Delta\vec\sigma}{\sqrt{2}}=
  \begin{bmatrix}
    \Delta^{3}/\sqrt{2}&\left(\Delta^{1}-i\Delta^{2}\right)/\sqrt{2}\\
    \left(\Delta^{1}+i\Delta^{2}\right)/\sqrt{2}&-\Delta^{3}/\sqrt{2}
  \end{bmatrix}&\equiv&
  \begin{bmatrix}
    \delta^{+}/\sqrt{2}&\delta^{++}\\
    \delta^{0}&-\delta^{+}/\sqrt{2}
  \end{bmatrix},\nonumber\\ 
  &&\delta^{0}=\frac{1}{\sqrt{2}}\left(v_{\Delta}+\delta+i\eta\right).
\end{eqnarray}
Here $\vec\sigma$ are the Pauli matrices.

The scalar sector kinetic terms are
\begin{equation}
  \label{eq:kinetic}
  {\cal L}_{\rm kinetic}=\left|D_{\mu}\Phi\right|^{2}
  +{\rm Tr}\left[\left(D_{\mu}\Delta\right)^{\dagger}\left(D_{\mu}\Delta\right)\right],
\end{equation}
where
\begin{eqnarray}
  D_{\mu}\Phi&=&\partial_{\mu}\Phi-i\frac{g}{2}A_{\mu}^{a}\sigma^{a}\Phi-i\frac{g'}{2}B_{\mu}\Phi,\\
  D_{\mu}\Delta&=&\left[\partial_{\mu}\Delta^{a}+g\varepsilon^{abc}A_{\mu}^{b}\Delta^{c}-ig'B_{\mu}\Delta^{a}\right]\frac{\sigma^{a}}{\sqrt{2}}=\nonumber\\
  &=& \partial_{\mu}\Delta-i\frac{g}{2}\left[A_{\mu}^{a}\sigma^{a},\Delta\right]-ig'B_{\mu}\Delta.
\end{eqnarray}
Hypercharge $Y_{\Phi}=1$ was substituted for isodoublet and
$Y_{\Delta}=2$ for isotriplet. The terms quadratic in vector boson
fields are the following: 
\begin{equation}
  \label{eq:VVHH}
  {\cal L}_{V^{2}}=g^{2}\left|\delta^{0}\right|^{2}W^{+}W^{-}
  +\frac{1}{2}g^{2}\left|\Phi^{0}\right|^{2}W^{+}W^{-}
  +\bar g^{2}\left|\delta^{0}\right|^{2}Z^{2}
  +\frac{1}{4}\bar g^{2}\left|\Phi^{0}\right|^{2}Z^{2}.  
\end{equation}

Vector boson masses are
\begin{equation}
  \label{eq:WZmass}
  \left\{
  \begin{array}{rcl}
    M_{W}^{2}&=&\frac{g^{2}}{4}\left(v^{2}+2v_{\Delta}^{2}\right),\\
    M_{Z}^{2}&=&\frac{\bar g^{2}}{4}\left(v^{2}+4v_{\Delta}^{2}\right).
  \end{array}\right.
\end{equation}

For the ratio of vector boson masses neglecting the radiative
corrections from isotriplet (not a bad approximation as far as the
heavy triplet decouples) we get:
\begin{equation}
  \label{eq:rho2}
  \frac{M_{W}}{M_{Z}}\approx\left(\frac{M_{W}}{M_{Z}}\right)_{\rm SM}\left(1-\frac{v_{\Delta}^{2}}{v^{2}}\right).
\end{equation}

Comparing the result of SM fit \cite[p.145]{PDG}, $M_{W}^{\rm
  SM}=80.381~\mbox{GeV}$, with the experimental value, $M_{W}^{\rm
  exp}=80.385(15)~\mbox{GeV}$, at $3\sigma$ level we get the following
upper bound:
\begin{equation}
  \label{eq:bound}
  v_{\Delta}<5~\mbox{GeV},
\end{equation}
and since the cross sections we are interested in are proportional to
$\left(v_{\Delta}\right)^{2}$ we will use an upper bound
$v_{\Delta}=5~\mbox{GeV}$ for numerical estimates in this section.

From the numerical value of Fermi coupling constant in muon decay we
obtain: 
\begin{equation}
  \label{eq:Fermi}
  v^{2}+2v_{\Delta}^{2}=\left(246~\mbox{GeV}\right)^{2},
\end{equation}
so for $v_{\Delta}\lesssim 5~\mbox{GeV}$ the value $v=246~\mbox{GeV}$
can be safely used in deriving (\ref{eq:bound}).

The scalar potential looks like:
\begin{eqnarray}
  \label{eq:potential}
  V(\Phi,\Delta)&=&
  -\frac{1}{2}m_{\Phi}^{2}\left(\Phi^{\dagger}\Phi\right)
  +\frac{\lambda}{2}\left(\Phi^{\dagger}\Phi\right)^{2}+\nonumber\\
  &+&M_{\Delta}^{2}{\rm Tr}\left[\Delta^{\dagger}\Delta\right]
  +\frac{\mu}{\sqrt{2}}\left(\Phi^{T}i\sigma^{2}\Delta^{\dagger}\Phi+h.c.\right),
\end{eqnarray}
which is a truncated version of the most general renormalizable
potential (see for example \cite{Dev}, eq. (2.6)). We may simply
suppose that the coupling constants which multiply the omitted terms
in the potential ($\lambda_{1},\lambda_{2},\lambda_{4}$, and
$\lambda_{5}$) are small. In the case of SM only the first line in
(\ref{eq:potential}) remains; mass of the Higgs boson equals
$m_{\Phi}=125~\mbox{GeV}$ while its expectation value
$v^{2}\approx m_{\Phi}^{2}/\lambda\approx\left(246~\mbox{GeV}\right)^{2}$,
$\lambda\approx 0.25$.

Since at the minimum of (\ref{eq:potential}) the following equations
are valid:
\begin{equation}
  \label{eq:minpot}
  \left\{
    \begin{array}{ccl}
      \frac{1}{2}m_{\Phi}^{2}&=&\frac{1}{2}\lambda v^{2}-\mu v_{\Delta},\\
      M_{\Delta}^{2}&=&\frac{1}{2}\mu\frac{\displaystyle v^{2}}{\displaystyle v_{\Delta}},
    \end{array}\right.
\end{equation}
for vev's of isodoublet and isotriplet we obtain:
\begin{eqnarray}
  \label{eq:v}
  v^{2}&=&\frac{m_{\Phi}^{2}M_{\Delta}^{2}}{\lambda
    M_{\Delta}^{2}-\mu^{2}},\\
  \label{eq:vdelta}
  v_{\Delta}&=&\frac{\mu m_{\Phi}^{2}}{2\lambda M_{\Delta}^{2}-2\mu^{2}}=\frac{\mu}{2}\frac{v^{2}}{M_{\Delta}^{2}}.
\end{eqnarray}

Quadratic in $\varphi,~\delta$ terms according to (\ref{eq:potential})
are
\begin{equation}
  \label{eq:HDmassterms}
  V(\varphi,\delta)=\frac{1}{2}m_{\Phi}^{2}\varphi^{2}
  +\frac{1}{2}M_{\Delta}^{2}\delta^{2}
  -\mu v\varphi\delta.
\end{equation}
Here and below the terms suppressed as $\left(v_{\Delta}/v\right)^{2}$
are omitted.

Denoting the states with the definite masses as $h$ and $H$ we obtain:
\begin{equation}
  \begin{bmatrix}
    \varphi\\
    \delta
  \end{bmatrix}  =
  \begin{bmatrix}
    \cos\alpha & -\sin\alpha\\
    \sin\alpha& \cos\alpha
  \end{bmatrix}
  \begin{bmatrix}
    h\\
    H
  \end{bmatrix},~ \tan 2\alpha=\frac{2\mu v}{M_{\Delta}^{2}-m_{\Phi}^{2}},
  \label{eq:Hhmix}
\end{equation}
\begin{eqnarray}
  M_{h}^{2}&=&
  \frac{1}{2}\left(m_{\Phi}^{2}+M_{\Delta}^{2}-\sqrt{\left(M_{\Delta}^{2}
        -m_{\Phi}^{2}\right)^{2}+4\mu^{2}v^{2}}\right)\approx m_{\Phi}^{2},\\
  M_{H}^{2}&=&
  \frac{1}{2}\left(m_{\Phi}^{2}+M_{\Delta}^{2}+\sqrt{\left(M_{\Delta}^{2}
        -m_{\Phi}^{2}\right)^{2}+4\mu^{2}v^{2}}\right)\approx M_{\Delta}^{2}.
  \label{eq:MH}
\end{eqnarray}
Since $\tan 2\alpha\approx 4v_{\Delta}/v\ll 1$, mass eigenstate $h$
consists mostly of $\varphi$ and $H$ consists mostly of $\delta$. We
suppose that the particle observed by ATLAS and CMS is $h$, so $M_{h}$
is about $125~\mbox{GeV}$. 

The scalar sector of the model in addition to the massless goldstone bosons,
which are eaten up by the vector gauge bosons, contains one double charged 
field $H^{++}$, one single charged field $H^{+}$, and three real
neutral fields $A,~H$, and $h$. $H^{+}$ is mostly $\delta^{+}$ with
small $\Phi^{+}$ admixture, $A$ is mostly $\eta$ with small $\chi$
admixture. All these particles except $h$ are heavy; their masses
equal $M_{\Delta}$ with small corrections proportional to
$v_{\Delta}^{2}/M_{\Delta}$. 

\subsection{$H$ decays}
\label{sec:Hdecays}

The second and fourth terms in potential (\ref{eq:potential})
contribute to $H\to 2h$ decays:
\begin{eqnarray}
  \label{eq:L3}
  \frac{\lambda}{2}\left(\Phi^{\dagger}\Phi\right)^{2}&\to&\frac{\lambda
    v}{2}\varphi^{3},\\
  \label{eq:ffd}
  \frac{\mu}{\sqrt{2}}\left(\Phi^{T}i\sigma^{2}\Delta^{\dagger}\Phi+h.c.\right)&\to&
  -\frac{\mu}{2}\delta\left(\varphi^{2}-\chi^{2}\right),
\end{eqnarray}
where in the second line $\chi$ is dominantly a goldstone state which
forms the longitudinal $Z$ polarization.

With the help of (\ref{eq:Hhmix}) we obtain the expression for the
effective lagrangian which describes $H\to 2h$ decay: 
\begin{equation}
  \label{eq:LHhh}
  {\cal L}_{Hhh}=
  \frac{\mu}{2}\left[1+\frac{3}{\left(\frac{M_{H}}{M_{h}}\right)^{2}-1}\right]Hh^{2}
  =v_{\Delta}\frac{M_{H}^{2}}{v^{2}}\left[1+\frac{3}{\left(\frac{M_{H}}{M_{h}}\right)^{2}-1}\right]Hh^{2}.
\end{equation}

In the see-saw type II model neutrino masses are generated by the Yukawa
couplings of isotriplet $\Delta$ with lepton doublets. These couplings
generate $H\to\nu\nu$ decays as well. As it was noted in \cite{Yagyu1}
for $v_{\Delta}>10^{-3}~\mbox{GeV}$ diboson decays dominate. It
happens because the amplitude of diboson decay is proportional to
$v_{\Delta}$, while Yukawa couplings $f_{i}$ are inversely
proportional to it, $f\sim m_{\nu}/v_{\Delta}$. That is why for
$v_{\Delta}\gtrsim 1~\mbox{GeV}$ leptonic decays are completely
negligible.

The amplitudes of $H\to ZZ$ and $H\to W^{+}W^{-}$ decays are contained in
(\ref{eq:VVHH}):
\begin{eqnarray}
  \label{eq:LHVV}
  {\cal L}_{HVV}&=&
  g^{2}\left(v_{\Delta}\cos\alpha-\frac{1}{2}v\sin\alpha\right)W^{+}W^{-}H
  +\bar g^{2}\left(v_{\Delta}\cos\alpha-\frac{1}{4}v\sin\alpha\right)Z^{2}H\nonumber\\
  &\approx& -g^{2}\frac{M_{h}^{2}/M_{H}^{2}}{1-M_{h}^{2}/M_{H}^{2}}v_{\Delta}W^{+}W^{-}H
  +\frac{\bar g^{2}}{2} \frac{1-2M_{h}^{2}/M_{H}^{2}}{1-M_{h}^{2}/M_{H}^{2}}v_{\Delta}Z^{2}H,
\end{eqnarray}
and we see that $H\to W^{+}W^{-}$ decay is suppressed (see, for example,
\cite{Perez}).

$H\to t\bar t$ decay occur through $\varphi$ admixture:
\begin{equation}
  \label{eq:LHtt}
  {\cal L}_{Ht\bar t}=\sin\alpha\frac{m_{t}}{v}t\bar tH
  =\frac{2v_{\Delta}/v}{1-M_{h}^{2}/M_{H}^{2}}\frac{m_{t}}{v}t\bar tH,
\end{equation}
as well as $H$ decay into two gluons:
\begin{equation}
  \label{eq:LHgg}
  {\cal L}_{Hgg}=\frac{\alpha_{s}}{12\pi}\sin\alpha G_{\mu\nu}^{2}.
\end{equation}
Let us note that all the amplitudes of $H$ decays are proportional to
triplet vev $v_{\Delta}$.

For the decay probabilities we obtain:
\begin{eqnarray}
  \label{eq:GHhh}
  \Gamma_{H\to hh}&=&\frac{v_{\Delta}^{2}}{v^{4}}\frac{M_{H}^{3}}{8\pi}
  \left[\frac{1+2\left(\frac{M_{h}}{M_{H}}\right)^{2}}{1-\left(\frac{M_{h}}{M_{H}}\right)^{2}}\right]^{2}
  \sqrt{1-4\frac{M_{h}^{2}}{M_{H}^{2}}},\\
  \Gamma_{H\to ZZ}&=&\frac{v_{\Delta}^{2}}{v^{4}}\frac{M_{H}^{3}}{8\pi}
  \left[\frac{1-2\left(\frac{M_{h}}{M_{H}}\right)^{2}}{1-\left(\frac{M_{h}}{M_{H}}\right)^{2}}\right]^{2}
  \left(1-4\frac{M_{Z}^{2}}{M_{H}^{2}}+12\frac{M_{Z}^{4}}{M_{H}^{4}}\right)
  \sqrt{1-4\frac{M_{Z}^{2}}{M_{H}^{2}}},\\
  \Gamma_{H\to WW}&=&\frac{v_{\Delta}^{2}}{v^{4}}\frac{M_{H}^{3}}{4\pi}
  \left[\frac{{M_{h}^{2}}/{M_{H}^{2}}}{1-\left(\frac{M_{h}}{M_{H}}\right)^{2}}\right]^{2}
  \left(1-4\frac{M_{W}^{2}}{M_{H}^{2}}+12\frac{M_{W}^{4}}{M_{H}^{4}}\right)
  \sqrt{1-4\frac{M_{W}^{2}}{M_{H}^{2}}},\\
  \Gamma_{H\to t\bar t}&=&\frac{v_{\Delta}^{2}}{v^{4}}\frac{N_{c}m_{t}^{2}M_{H}}{2\pi}
  \frac{1}{\left(1-M_{h}^{2}/M_{H}^{2}\right)^{2}}\left(1-4\frac{m_{t}^{2}}{M_{H}^{2}}\right)^{3/2},
\end{eqnarray}
where $N_{c}=3$ is the number of colors. Finally for the width of
decay into two gluon jets we obtain:
\begin{equation}
  \label{eq:GHgg}
  \Gamma_{H\to gg}=\frac{v_{\Delta}^{2}}{v^{4}}\frac{M_{H}^{3}}{2\pi}
  \left(\frac{\alpha_{s}}{3\pi}\right)^{2}\left(1-\frac{M_{h}^{2}}{M_{H}^{2}}\right)^{-2},
\end{equation}
and it is always negligible.

In what follows we suppose that $M_{H}<350~\mbox{GeV}$ and the decay
$H\to t\bar t$ is forbidden kinematically. Let us note that even for 
$M_{H}>350~\mbox{GeV}$ the branching ratio of $H\to 2h$ decay is
large, however $H$ production cross section becomes small due to 
the large $H$ mass.

The lighter $H$ the larger its production cross section, however, for
$M_{H}<250~\mbox{GeV}$ the decay $H\to 2h$ is kinematically
forbidden. That is why for numerical estimates we took the value
$M_{H}=300~\mbox{GeV}$ for which $H\to 2h$ and $H\to ZZ$ decays
dominate\footnote{The decay $H\to ZZ\to
  \left(l^{+}l^{-}\right)\left(l^{+}l^{-}\right)$ provides great  
  opportunity for the discovery of heavy Higgs $H$.} and $\Gamma_{H\to
  2h}/\Gamma_{H\to ZZ}\approx 4$. Thus $300~\mbox{GeV}$ (or a little
bit lighter) $H$ mostly decays to two $125~\mbox{GeV}$ Higgs bosons. 

A technical remark: the equality $\Gamma_{H\to hh}=\Gamma_{H\to ZZ}$ in
the limit $M_{H}\gg M_{h},M_{H}\gg M_{Z}$ follows from the equality
(up to the sign) of $H\to 2h$ and $H\to 2\chi$ decay amplitudes, see
(\ref{eq:ffd}).

\subsection{$H$ production at LHC}
\label{sec:production}

The dominant mechanism of $H$ production is the gluon fusion, cross
section of which equals that of SM Higgs production multiplied by
$\sin^{2}\alpha\approx
\left[\left(2v_{\Delta}/v\right)/\left(1-M_{h}^{2}/M_{H}^{2}\right)\right]^{2}\approx
2.4\cdot 10^{-3}$. In Table \ref{tab:gg} the relevant numbers are
presented. All the numbers correspond to $14~\mbox{TeV}$ LHC energy.

\begin{table}[t]
  \caption{The cross sections of Higgs production via $gg$
    fusion. Values for the SM Higgs are taken from Table 4 in
    \cite{Handbook}. All numbers in this and following tables
    correspond to $14~\mbox{TeV}$ LHC energy.}
  \label{tab:gg}
  \centering
  \begin{tabular}[t]{|>{\centering\arraybackslash}p{0.2\textwidth}|>{\centering\arraybackslash}p{0.2\textwidth}|>{\centering\arraybackslash}p{0.2\textwidth}|}
    \hline
    $M_{h}~\left(\mbox{GeV}\right)$ & \centering 125 & 300\\
    \hline
    $\sigma_{gg\to h}~\left(\mbox{pb}\right)$ & $49.97\pm 10\%$ & $11.07\pm 10\%$\\
    \hline
    \hline
    $M_{H}~\left(\mbox{GeV}\right)$ & X & 300\\
    \hline
    $\sigma_{gg\to H}~\left(\mbox{fb}\right)$ & X & $25\pm 10\%$\\
    \hline
  \end{tabular}
\end{table}

The subdominant mechanisms of $H$ production are $ZZ$ fusion and
associative $ZH$ production. Comparing $ZZh$ and $ZZH$ vertices we
will recalculate the cross sections of SM processes of $h$ production into
that of $H$ production. In SM we have
\begin{equation}
  \label{eq:LhZZ}
  {\cal L}_{hZZ}=\frac{1}{4}\bar g^{2}vZ^{2}h.
\end{equation}

From (\ref{eq:LHVV}) we get:
\begin{equation}
  \label{eq:sZZH}
  \sigma_{ZZ\to H}=\left(\frac{2v_{\Delta}}{v}\frac{1-2M_{h}^{2}/M_{H}^{2}}{1-M_{h}^{2}/M_{H}^{2}}\right)^{2}
  \times\left(\sigma_{ZZ\to h}\right)^{\rm SM}
  \approx 10^{-3}\times\left(\sigma_{ZZ\to h}\right)^{\rm SM},
\end{equation}
the same relation holds for $Z^{*}\to ZH$ associative production cross
section.

We separate VBF cross section of SM Higgs production into that in
$W^{+}W^{-}$ fusion (which dominates) and in $ZZ$ fusion (which is the
one that matters for $H$ production) with the help of the computer code HAWK
\cite{HAWK}. The obtained results are presented in Table \ref{tab:vbf}.

In Table \ref{tab:zh} the results for the associative $ZH$ production cross
sections are presented.

\begin{table}[t]
  \caption{The cross sections (QCD NLO) of scalar bosons production in VBF
    calculated with the help of HAWK (see also Table 10 in \cite{Handbook}).} 
  \label{tab:vbf}
  \centering
  \begin{tabular}[t]{|>{\centering\arraybackslash}p{0.2\textwidth}|>{\centering\arraybackslash}p{0.2\textwidth}|>{\centering\arraybackslash}p{0.2\textwidth}|}
    \hline
    $M_{h}~\left(\mbox{GeV}\right)$ & 125 & 300\\
    \hline
    $\sigma_{VV\to h}~\left(\mbox{fb}\right)$& 4342(5) & 1418(1)\\
    \hline
    $\sigma_{W^{+}W^{-}\to h}~\left(\mbox{fb}\right)$& 3272(4) & 1053(1)\\
    \hline
    $\sigma_{ZZ\to h}~\left(\mbox{fb}\right)$& 1087(1) & 365(1) \\
    \hline
    \hline
    $M_{H}~\left(\mbox{GeV}\right)$ & X & 300\\
    \hline
    $\sigma_{ZZ\to H}~\left(\mbox{fb}\right)$& X & 0.365(1)\\
    \hline
  \end{tabular}
\end{table}

\begin{table}[t]
  \caption{The cross sections of the associative SM Higgs production from
    Table 14 in \cite{Handbook} and of associative $H$ production recalculated
    with the help of (\ref{eq:sZZH}).}
  \label{tab:zh}
  \centering
  \begin{tabular}[t]{|>{\centering\arraybackslash}p{0.2\textwidth}|>{\centering\arraybackslash}p{0.2\textwidth}|>{\centering\arraybackslash}p{0.2\textwidth}|}
    \hline
    $M_{h}~\left(\mbox{GeV}\right)$ & 125 & 300\\
    \hline
    $\sigma_{W^{*}\to Wh}~\left(\mbox{fb}\right)$& $1504\pm4\%$ & $67.6\pm4\%$\\
    \hline
    $\sigma_{Z^{*}\to Zh}~\left(\mbox{fb}\right)$& $883\pm5\%$ & $41.6\pm5\%$\\
    \hline
    \hline
    $M_{H}~\left(\mbox{GeV}\right)$ & X & 300\\
    \hline
    $\sigma_{Z^{*}\to ZH}~\left(\mbox{fb}\right)$& X & $0.0416\pm5\%$\\
    \hline
  \end{tabular}
\end{table}

We see that gluon fusion dominates $H$ production at LHC. Using
model parameters $v_{\Delta}= 5~\mbox{GeV}$ and
$M_{H}=300~\mbox{GeV}$, we obtain that the branching ratio of $H\to
2h$ decay equals $\approx 80\%$. Thus, decays of $H$ provide $\approx
20~\mbox{fb}$ of double $h$ production cross section in addition to
$40~\mbox{fb}$ coming from SM. However, unlike SM in which $2h$
invariant mass is spread along rather large interval, in the case of
$H$ decays $2h$ invariant mass equals $M_{H}$.

\section{$H$ production enhancement in Georgi--Machacek variant of see-saw type II model}
\label{sec:GM}

The amplitudes of $H$ production both via $gg$ fusion and VBF are
proportional to the triplet vev $v_{\Delta}$ and due to the upper
bound $v_{\Delta}<5~\mbox{GeV}$ these amplitudes and the corresponding
cross sections are severely suppressed. 

The triplet vev $v_{\Delta}$ should be small in order to avoid the 
noticeable violation of custodial symmetry which guarantees the
degeneracy of $W$ and $Z$ bosons in the SM at tree level in the limit
$g'=0,~\cos\theta_{W}=1$. The vacuum expectation value of the complex
isotriplet $\vec\Delta$ with hypercharge $Y_{\Delta}=2$ violates the custodial
symmetry, see (\ref{eq:WZmass}). The custodial symmetry is preserved
when two isotriplets (complex $\vec\Delta$ and real $\vec\xi$ with
$Y_{\xi}=0$) are added to SM and when vev's of their neutral components are
equal \cite{Georgi}. Thus in GM variant of see-saw type II model
$v_{\Delta}$ is not bounded by (\ref{eq:bound}) and can be
considerably larger. Instead of (\ref{eq:WZmass}) in GM model we have:
\begin{equation}
  \label{eq:WZmass_GM}
  \left\{
  \begin{array}{rcl}
    M_{W}^{2}&=&\frac{g^{2}}{4}\left(v^{2}+4v_{\Delta}^{2}\right),\\
    M_{Z}^{2}&=&\frac{\bar g^{2}}{4}\left(v^{2}+4v_{\Delta}^{2}\right),
  \end{array}\right.
\end{equation}
and instead of (\ref{eq:Fermi}):
\begin{equation}
  \label{eq:Fermi_GM}
  v^{2}+4v_{\Delta}^{2}=\left(246~\mbox{GeV}\right)^{2}.
\end{equation}

Note that our $v_{\Delta}$ is by $\sqrt{2}$ bigger than what is
usually used in the papers devoted to GM model; our $v$ is also
usually denoted by $v_{\Phi}$, while the value $246~\mbox{GeV}$ is
denoted by $v$.

The scalar particles are conveniently classified in GM model by their
transformation properties under the custodial $SU(2)$. Two singlets
which mix to form mass eigenstates $h$ and $H$ are:
\begin{equation}
  \label{eq:H10H20}
  \left\{
  \begin{array}{rcl}
    H_{1}^{0}&=&\varphi,\\
    H_{2}^{0}&=&\sqrt{\frac{2}{3}}\delta+\sqrt{\frac{1}{3}}\xi^{0},
  \end{array}\right.
\end{equation}
see, for example, \cite{Logan2014}. Due to considerable admixture of
$\xi^{0}$ in $H_{2}^{0}$ the $HW^{+}W^{-}$ coupling constant is not
suppressed and three modes of $H$ decays are essential: $H\to hh,
~H\to W^{+}W^{-},~H\to ZZ$.

The recently discovered Higgs boson should be identified with
$h$. The deviations of $h$ couplings to vector bosons and fermions from
their values in SM lead to the upper bound on $v_{\Delta}$. These
deviations in the limit of heavy scalar triplets were studied in a
recent paper \cite{Logan2014} (see also \cite{Englert2013}). From equations (59) and (61) of
\cite{Logan2014} we get the following estimates for the ratios of the
$hVV$ (here $V=W,~Z$) and $h\bar ff$ coupling constants to that in SM:
\begin{equation}
  \label{eq:couplings_GM}
  \left\{
  \begin{array}{rcl}
    k_{V}&\approx& 1+3\left(\frac{v_{\Delta}}{v}\right)^{2},\\
    k_{f}&\approx& 1-\left(\frac{v_{\Delta}}{v}\right)^{2}.
  \end{array}\right.
\end{equation}

Since at LHC the Higgs boson $h$ is produced mainly in gluon fusion
through $t$-quark triangle, for the ratio of the cross sections to that
in SM we get:
\begin{equation}
  \label{eq:mu_GM}
  \left\{
  \begin{array}{rcl}
    \mu_{\tau\bar\tau}&\approx& 1-\left(2\frac{v_{\Delta}}{v}\right)^{2},\\
    \mu_{VV}&\approx& 1+\left(2\frac{v_{\Delta}}{v}\right)^{2}.
  \end{array}\right.
\end{equation}

Since $h\to b\bar b$ decay is studied in associative production,
$V^{*}\to Vh\to Vb\bar b$, we get
\begin{equation}
  \label{eq:mubb_GM}
  \mu_{b\bar b}\approx 1+\left(\frac{2v_{\Delta}}{v}\right)^{2}.
\end{equation}

Finally in case of $h\to\gamma\gamma$ decay SM factor $16/9-7$ in the
amplitude is modified in the following way:
\begin{eqnarray}
  \frac{16}{9}-7&\to&
  \left[1-\left(\frac{v_{\Delta}}{v}\right)^{2}\right]
  \left[\frac{16}{9}\left(1-\left(\frac{v_{\Delta}}{v}\right)^{2}\right)
  -7\left(1+3\left(\frac{v_{\Delta}}{v}\right)^{2}\right)\right]=\nonumber\\
&=&\frac{16}{9}\left(1-2\left(\frac{v_{\Delta}}{v}\right)^{2}\right)
  -7\left(1+2\left(\frac{v_{\Delta}}{v}\right)^{2}\right),
  \label{eq:hgg_GM}
\end{eqnarray}
where the first factor in the first line takes into account damping of
$h$ production in gluon fusion.\footnote{We take into account only
  $t$-quark and $W$-boson loops omitting the loops with charged
  Higgses.}

Let us suppose that $v_{\Delta}$ is ten times larger than the number
used in Section \ref{sec:seesaw},
$v_{\Delta}^{GM}=50~\mbox{GeV}$. Then from (\ref{eq:Fermi_GM}) we get
$v^{GM}\approx 225~\mbox{GeV}$, and $\mu_{\tau\bar\tau}\approx 0.8$,
while $\mu_{WW}=\mu_{ZZ}=\mu_{bb}\approx 1.2$. From (\ref{eq:hgg_GM})
we get: $\mu_{\gamma\gamma}\approx 1.4$.  With the up-to-date level of
the experimental accuracy one can not exclude these deviations of the
quantities $\mu_{i}$ from their SM values
$\left(\mu_{i}\right)^{SM}\equiv 1$.

One order of magnitude growth of $v_{\Delta}$ leads to two orders of
magnitude growth of $H$ production cross section. Hence $300~\mbox{GeV}$
heavy Higgs boson $H$ can be produced at $14~\mbox{TeV}$ LHC with
$2~\mbox{pb}$ cross section which should be large enough for it to be
discovered prior to HL-LHC. The search strategy should be the same as
for the SM Higgs boson: $gg\to H\to ZZ$ decay is a golden discovery mode,
the cross section of which can be as large as $\left(2~\mbox{pb}\right)\times{\rm
  Br}\left(H\to ZZ\right)^{\rm GM}$, where ${\rm Br}\left(H\to
  ZZ\right)^{\rm GM}$ depends on the model parameters, see \cite{Logan2014}.

\section{Conclusions}
\label{sec:conclusions}

The case of extra isotriplet(s) provides rich Higgs sector
phenomenology with additional to SM Higgs boson charged and neutral
scalar particles. With the growth of triplet vev, production cross
section of new scalar grows and the dominant decays of new
particles become decays to gauge and lighter scalar bosons. The charged
scalars ($\Phi^{++},~\Phi^{+}$) are produced through electroweak
interactions. The bounds on the model parameters from nondiscovery of
$\Phi^{++}$ and $\Phi^{+}$ with the $8~\mbox{TeV}$ LHC data and
the prospects of their discovery at $14~\mbox{TeV}$ LHC are discussed in
particular in \cite{Yagyu2}. In the present paper we have discussed
the neutral heavy Higgs production at LHC in which the gluon fusion
dominates. $H\to 2h$ decay contributes significantly to the double
Higgs production and even may dominate in the GM variant of the
see-saw type II model. The best discovery mode for $H$ is the ``golden
mode'' $pp\to HX\to ZZX$, and its cross section can be only few times
smaller than for the heavy SM Higgs.

After this paper had been written, paper \cite{BhCh2014} appeared in
arXiv in which the enhancement of double Higgs production due to heavy
Higgs decay is considered in the framework of MSSM model with two
isodoublets. $H\to 2h$ resonant decay in MSSM at small $\tan\beta$ was
previously analyzed in \cite{Englert2012_2}.

We are grateful to A. Denner for the clarifications concerning HAWK
code and to C. Englert and J. Zurita for providing us with
relevant references. The authors are partially supported under the grants
No. 12-02-00193, 14-02-00995, and NSh-3830.2014.2. SG was also
supported by Dynasty Foundation and by the Russian Federation
Government under grant No. 11.G34.31.0047.

\end{document}